\documentclass{aa}
\usepackage{graphicx}
\usepackage{color}

\def\gtrsim{\mathrel{\hbox{\rlap{\hbox{\lower4pt\hbox{$\sim$}}}\hbox{$>$}}}}
\def\ltsim{\mathrel{\hbox{\rlap{\hbox{\lower4pt\hbox{$\sim$}}}\hbox{$<$}}}}

\begin{document}

\title{ 
The magnetic field of Betelgeuse: a local dynamo from giant convection cells?
\thanks{Based on observations  obtained at the T\'elescope Bernard Lyot (TBL) at Observatoire du Pic du Midi, CNRS/INSU and Universit\'e de Toulouse, France.}}

\author{M. Auri\`ere\inst{1}, J.-F. Donati\inst{1}, R. Konstantinova-Antova\inst{2,1}, G. Perrin\inst{3}, P. Petit\inst{1}, T. Roudier\inst{1}}
\offprints{M. Auri\`ere, {\tt michel.auriere@ast.obs-mip.fr}}

\institute{Laboratoire d'Astrophysique de Toulouse- Tarbes, Universit\'e de Toulouse, CNRS, 57 Avenue d'Azereix, 65000 Tarbes, France
\email{michel.auriere@ast.obs-mip.fr, donati@ast.obs-mip.fr, petit@ast.obs-mip.fr, thierry.roudier@ast.obs-mip.fr}
\and
Institute of Astronomy, Bulgarian Academy of Sciences, 72 Tsarigradsko shose, 1784 Sofia, Bulgaria
\email{renada@astro.bas.bg}
\and
Observatoire de Paris, LESIA, UMR8109, 92190 Meudon, France
\email{guy.perrin@obspm.fr}}

 \date{Received ??; accepted ??}

\abstract
 {Betelgeuse is an M supergiant with a complex and extended atmosphere, which also harbors spots and giant granules at its surface. A possible magnetic field could contribute to the mass loss and to the heating of the outer atmosphere.}
{We observed Betelgeuse, to directly study and infer the nature of its magnetic field.}
{We used the new-generation spectropolarimeter NARVAL and the least square deconvolution (LSD) method to detect circular polarization within the photospheric absorption lines of Betelgeuse. }
{We have unambiguously detected a weak 
Stokes $V$ signal in the spectral lines of Betelgeuse, and measured the related surface-averaged longitudinal magnetic field $B_\ell$ at 6 different epochs over one month.  The detected longitudinal field is about one Gauss and is apparently increasing on the time scale of  our observations.}
{This work presents the first direct detection of the magnetic field of Betelgeuse. This magnetic field may be associated to the giant convection cells that could enable a ``local dynamo''.}

   \authorrunning {M. Auri\`ere et al.}
   \titlerunning {The magnetic field of Betelgeuse}

\maketitle

\section{Introduction}

Betelgeuse ($\alpha$ Ori, HD 39801) is a nearby M2Iab supergiant, and a star with the largest apparent diameter. It has naturally deserved direct imaging and interferometric studies of its surface, which was found to deviate from circular symmetry and to present variable behavior (Wilson et al. 1997, Haubois et al. 2009). The presence of a few spots and giant granules appeared as a natural explanation for the imaging and interferometric observations and their variability. A magnetic field at the surface of this star has also been suggested  by Dorch and Freytag (2003). Numerical simulations of convection now match interferometric observations and enable determination of the convection pattern on Betelgeuse (Chiavassa et al. 2010). 

 In this context, we have undertaken a very sensitive magnetic study of Betelgeuse, using the new generation spectropolarimeter NARVAL in operation at T\'elescope Bernard Lyot (TBL, Observatoire du Pic du Midi). We present here the result of a short campaign in March-April 2010. We report the definite detection of a weak magnetic field at the surface of Betelgeuse. Section 2 describes our observations and Sect. 3 our results. We discuss the nature of the magnetic field in Sect. 4 and give our conclusions in Sect. 5.

\section{Observations with NARVAL}

Observations of Betelgeuse were obtained at the TBL using the spectropolarimeter NARVAL, which is a twin of  ESPaDOnS (Donati et al. 2006). 

\begin{table}
\caption{Log of observations of Betelgeuse (for details, see Sect. 2).  }          

\label{table:1}   
\centering                         
\begin{tabular}{c c c c c}     
\hline\hline               
Date          &  HJD      &Tot Exp.& $B_\ell$ & $\sigma$  \\
              &(2 450 000 +)& s      & G        & G         \\
\hline                        
14 March 2010 & 5270.397  &320     &0.49      &0.16       \\
15 March 2010 & 5271.372  &320     &0.89      &0.15       \\
17 March 2010 & 5273.311  &256     &0.74      &0.21       \\
22 March 2010 & 5278.362  &316     &1.07      &0.15       \\
09 April 2010 & 5296.318  &320     &1.58      &0.14       \\
17 April 2010 & 5304.336  &272     &1.61      &0.19       \\
\hline
\hline                            
\end{tabular}

\end{table}

We observed Betelgeuse on 6 dates in March-April 2010. A standard circular polarization observation consists of a series of 4 sub exposures. 

To avoid saturation, we performed short exposures (3-5 s, sky-quality depending) for each sub exposure. We got 16 to 20 Stokes $V$ / Stokes $I$ series each night, which were averaged. Also included in the output are  ``diagnostic null" spectra $N$, which are in principle featureless, and therefore serve to diagnose the presence of spurious contributions to the Stokes $V$ spectrum. Each single spectrum used in this work has a peak signal-to-noise ratio (S/N) in Stokes $I$ per 2.6 km s$^{-1}$ spectral bin between 1700 and 2100. Details on the used procedure can be found in Donati et al. (1997) and Auri\`ere et al. (2009). Table 1 gives the log of observations with the dates,  the mean Heliocentric Julian Date (HJD) of the binned measurement of each night, and the total exposure time. To obtain a high-precision diagnosis of the spectral line circular polarization, the least-square 
deconvolution 
(LSD, Donati et al. 1997) was applied to each reduced Stokes $I$ and $V$ spectrum. We used a solar abundance line mask calculated from data provided by VALD (Kupka et al. 1999) for an effective 
temperature of 3750 K, $\log g =0.0$, and a microturbulence of 4.0 km s$^{-1}$, consistent with the physical parameters of Betelgeuse (Josselin and Plez 2007, Lambert et al. 1984). The mask contains about 15000 atomic lines with a central depth greater than 40\% of the continuum.

From these mean Stokes profiles, we computed the surface-averaged longitudinal magnetic field  $B_\ell$ in G, using the first-order moment method (Rees \& Semel 1979), adapted to LSD profiles (Donati et al. 1997, Wade et al. 2000). The measurements of $B_\ell$ are presented in Table 1 with their 1 $\sigma$ error in G. These errors are computed from photon statistical error bars propagated when reducing the polarization spectra and computing the LSD profiles (Donati et al. 1997).

\section{Results of the observations}
\subsection{Zeeman detection of a weak magnetic field on Betelgeuse}

Each night, a characteristic magnetic Stokes $V$ signature appears on the average of the obtained LSD profiles, giving a definitive detection when using the LSD statistical detection criteria (Donati et al. 1997). The upper part of Fig. 1 shows the $S/N^2$-weighted average of the 16 LSD profiles obtained on 15 March 2010 for Betelgeuse.  The averaged Stokes $V$ profile shows a definite Zeeman detection with a reduced $\chi^{2}$ equal to 3. The polarization signal is weak with an amplitude of about $3\times 10^{-5}$ of the continuum. Figure 2 shows the LSD-averaged Stokes $V$ and diagnostic $N$ profiles for the 6 dates of observations. Table 1 shows that the corresponding surface-averaged longitudinal magnetic field $B_\ell$ is about 1G. Even weaker Zeeman detections than in Betelgeuse were observed in Pollux (Auri\`ere et al. 2009) and Vega (Ligni\`eres et al. 2009). For comparison, the lower part of Fig.1 shows the averaged LSD profiles we obtained for the K5III giant Aldebaran the same night and for similar S/N (16 spectra were averaged): in this case, no Stokes $V$ signal was obtained.

\subsection{Possible variation of the magnetic field and associated measurements }

Table 1 shows an increase in the values of $B_\ell$ measured over the span of observations. The variation is at the 5 $\sigma$ level if the numbers for first and last observations are taken at face value. Figure 3 presents the associated plot of $B_\ell$ variations. In Fig. 2, the shape of the Stokes V profile appears to change between March and April. A month-scale variation of the magnetic field is therefore suggested, but it needs to be confirmed.

In addition, we measured  the radial velocity of Betelgeuse and investigated the CaII H\&K profiles for each night. The radial velocity $RV$ of Betelgeuse was measured from the averaged LSD Stokes $I$ profiles using a Gaussian fit. The long-term stability of NARVAL is about 30 m$s^{-1}$ (Auri\`ere et al. 2009), but the absolute uncertainty of individual measurements relative to the local standard of rest is about 1 km $s^{-1}$. In March, $RV$ was found to be constant, at about 24.75 km $s^{-1}$. On our April observations, $RV$ was found to decrease by  about 1 km $s^{-1}$. As to the CaII H\&K lines, in March our spectra looked similar to those presented by Toussaint and Reimers (1989) but with an opposite $V/R$ asymmetry (ratio between the violet (V) and red (R) components of the double-peaked emission line). Some decrease in the emission in CaII H\&K appears to occur in April compared to its level in March.

\begin{figure}
\centering
\includegraphics[width=7 cm,angle=0] {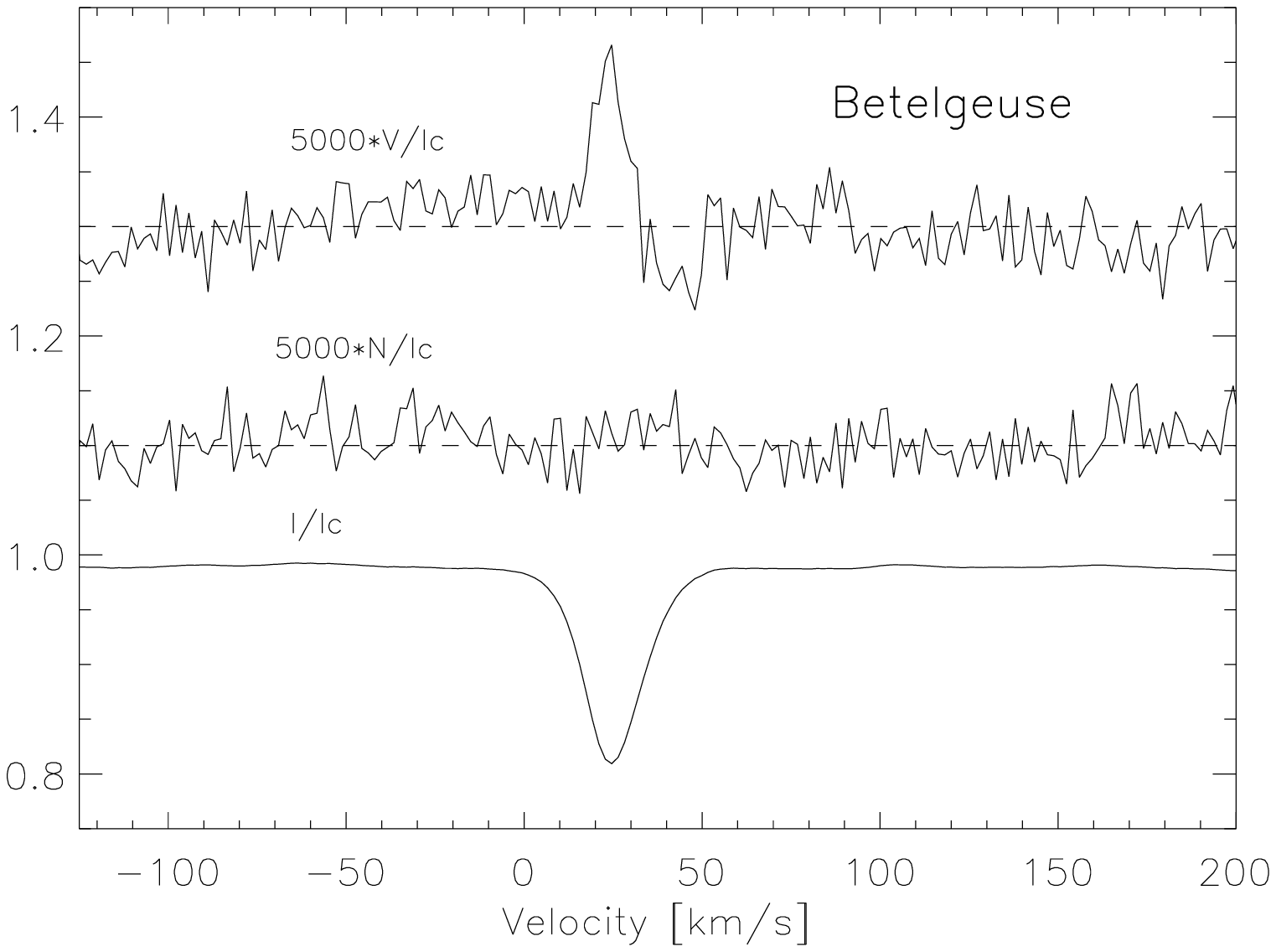} 

\includegraphics[width=7 cm,angle=0] {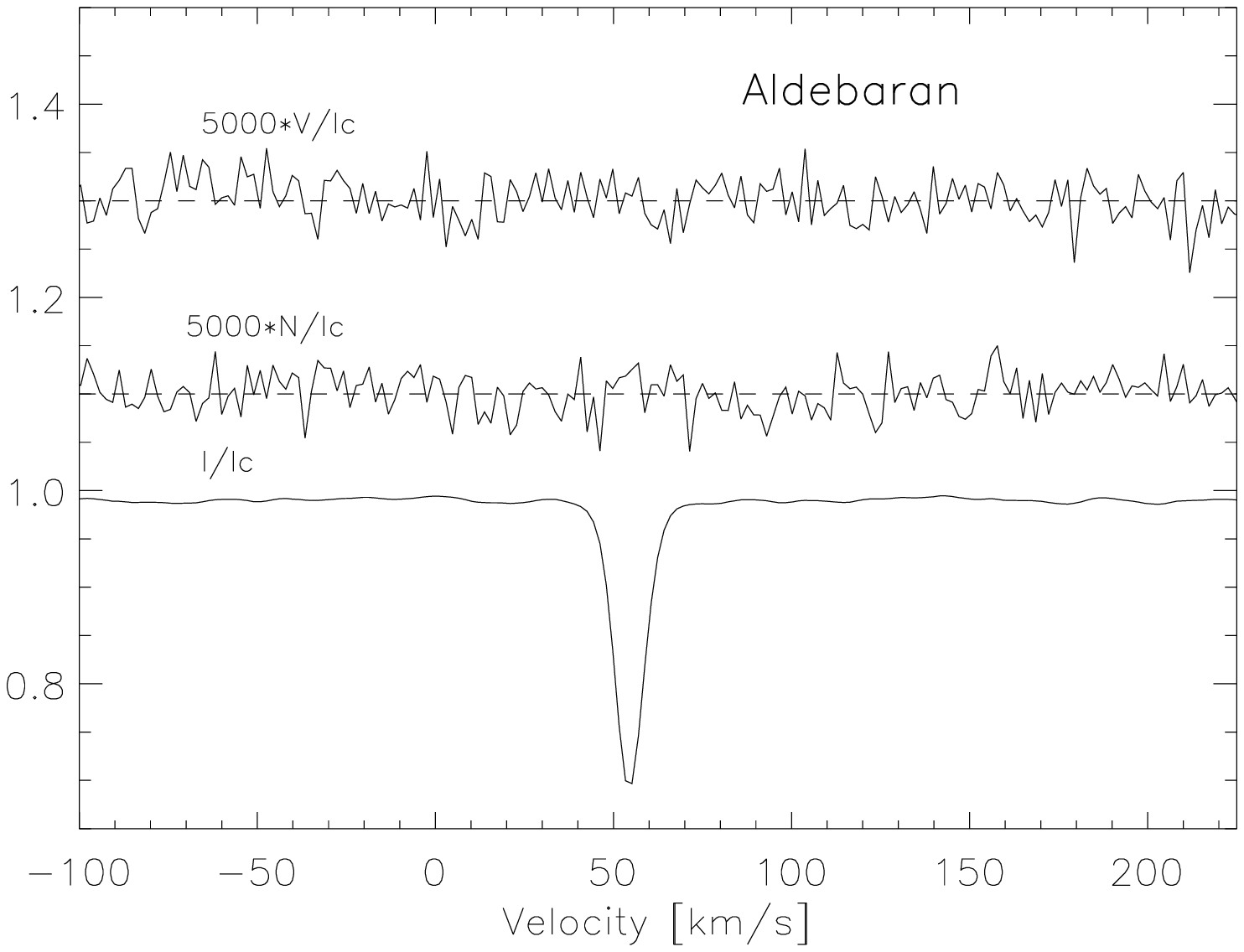} 

\caption{Mean LSD profiles of Betelgeuse (upper figure, definitive Zeeman detection) and Aldebaran (lower figure, no Zeeman detection) as obtained with NARVAL on 15 March 2010. For each figure and from top to bottom are: Stokes $V$, zero polarization $N$, and Stokes $I$ profiles. For display purposes, the profiles are shifted vertically, and the Stokes $V$ and diagnostic $N$ profiles are expanded by a factor of 5000. The dashed line illustrates the zero level for the Stokes $V$ and zero polarization $N$ profiles.}
\end{figure}

\begin{figure}
\centering

\includegraphics[width=7 cm,angle=0] {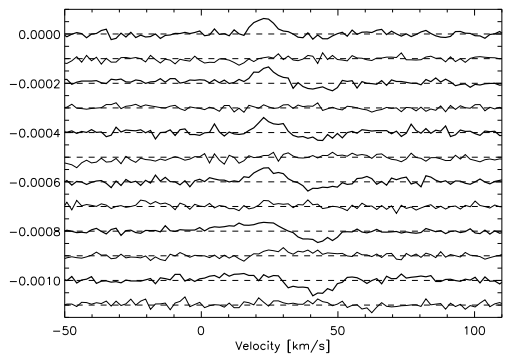}

\caption{Averaged Stokes $V$ (thick line) and zero polarization $N$ (thin line) profiles of Betelgeuse on each observed night of March-April 2010. From top to bottom, the 6 dates are those in Table 1. For display purposes, each profile is expanded by a factor of 2. The dashed lines illustrate the zero level for the Stokes $V$ and diagnostic $N$ profiles.}
\end{figure}

\begin{figure}
\centering

\includegraphics[width=7 cm,angle=0] {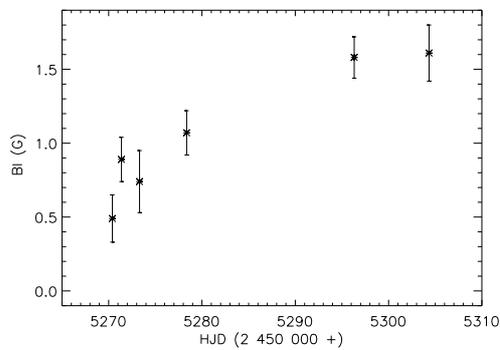}

\caption{Variations in the longitudinal magnetic field of Betelgeuse with HJD in March-April 2010. X-axis: HJD in days: 2 450 000 + . Y axis: $B_\ell$ in G (bars show 1 $\sigma$ errors).}
\end{figure}

\section {The magnetic field of Betelgeuse}

\subsection {The context of stellar magnetism}

The observational knowledge of the stellar magnetic fields has  improved tremendously these past years, in particular for main-sequence and pre-main-sequence stars (e.g. MAPP project, Donati et al. 2010) mostly thanks to the improvement in instrumental performances (Donati and Landstreet 2009). As to the evolved stars, strong magnetic fields (100 G or more) have been observed at the surfaces of very fast rotating giants, including RS CVn binaries or FK Com stars (e.g. HD 199178, Petit et al. 2004). Weaker magnetic fields ($B_\ell$ of a few G to some tens of G) have been detected with NARVAL or ESPaDOnS at the surface of active single giants, which rotate faster than the bulk of the red giant class (Konstantinova-Antova et al. 2008, 2010, L\`ebre et al. 2009). For all these active stars, the origin of the magnetic field is very likely a dynamo. Magnetic fields were also detected in slowly rotating giants, including EK Eri, which presents a strong magnetic field (Auri\`ere et al. 2008) and Pollux, which presents a very weak magnetic field (Auri\`ere et al. 2009). These Zeeman-detected evolved stars included only intermediate mass objects. 

Currently, an increasing number of massive stars have been detected as magnetic, in particular thanks to the systematic investigation of the MiMeS collaboration (e.g. Grunhut et al. 2009), which concerns main-sequence and pre-main sequence stars. An investigation of the magnetic fields of AFGKM supergiants is also under way and first Zeeman detections were presented recently (Grunhut et al. 2010). Betelgeuse appears to be the first M supergiant to be detected as magnetic.

\subsection {Activity of Betelgeuse and its magnetic field}

Betelgeuse is a highly variable star in a general sense, because it is an irregular pulsating variable star that presents a wide range of photometric and spectral variations (Goldberg 1984, Gray 2008). For the chromospheric lines, which can be tracers of magnetic activity, Ca II H\&K present variations  (Toussaint and Reimers, 1989), as well as MgII k (Dupree et al. 1987). No X-ray emission, which can be due to coronal heating, could be detected in any observation of Betelgeuse, and very weak upper limits in flux were reached (Maggio et al. 1990, Posson-Brown et al. 2006). The reason could be that the magnetic loops are ``buried'' in the highly extended chromospheric material, as suggested for cool giants by Ayres et al. (2003).

Because the rotational period of Betelgeuse is expected to last several years (2335 days: AAVSO data, Stothers and Leung 1971; 17 years: Uitenbroek et al. 1998) a classical solar-type dynamo is not expected to operate there. Taking its large radius into account (R=645 $R_{\odot}$, Perrin et al. 2004), the fossil field from a magnetic main sequence star would be too diluted to provide an efficient remnant as in EK Eri (G8III/IV, Stepie\'n 1993, Strassmeier et al. 1999, Auri\`ere et al. 2008).

Theoretical predictions (Schwarzschild 1975) and interferometric observations (e.g. Haubois et al. 2009, Chiavassa et al. 2010)
 suggest there are large convection cells on the surface of Betelgeuse. The 3D convection simulations of Betelgeuse have already been carried out (Freytag et al. 2002; Dorch 2004) and these studies also suggest there are large convective cells and furthermore that a magnetic field could be sustained. The Betelgeusian dynamo would belong to the class of so-called ``local small-scale dynamos'', though the generated magnetic field is both local and large scale (Dorch 2004). This kind of dynamo would work even without rotation (Freytag et al. 2002) and has the same nature as the local dynamo, possibly contributing to the small-scale magnetic field on the solar surface (cf. Cattaneo 1999). Dorch (2004) presents a detailed numerical MHD simulation of Betelgeuse that shows that magnetic spots of strength up to 500 G may exist but with a small filling factor. Our detection and measurement do prove the existence of a magnetic field on the surface of Betelgeuse. A month-scale variation may have been observed and has to be confirmed. Variation in the surface pattern on Betelgeuse has been observed on the same time scale with interferometry (Wilson et al. 1997). This time scale is unlikely to be linked to the long rotational period and might stem from a local intrinsic variability induced by a local dynamo.

\section{Conclusion}

We Zeeman-detected a magnetic field on the surface of Betelgeuse and measured its surface-averaged longitudinal component to be about 1 G. Because  M supergiants are very slow rotators and since they have very  extended envelopes, neither a solar-type dynamo nor a remnant of a fossil magnetic field from a magnetic main sequence progenitor is expected to be the origin of the magnetic field of Betelgeuse. A possible explanation for its existence can therefore be the giant convective cells predicted to exist on the surface of the star (Schwarzschild 1975), explaining the observations with a high angular resolution (e.g. Haubois et al. 2009, Chiavassa et al. 2010) 
and appearing in the numerical simulations (Freytag et al. 2002; Dorch 2004). Certainly, the established magnetic field plays a role in the mass loss and in the heating of the outer atmosphere of Betelgeuse (Lim et al. 1998).

\begin{acknowledgements}
 We acknowledge the use of the databases VALD (Vienna, Austria) and Simbad (CDS, Strasbourg, France). We thank  the TBL team for providing service observing with NARVAL. R. K.-A. acknowledges  the possibility of working for six months in 2010 as a visiting researcher in LATT, Tarbes under Bulgarian NSF grant DSAB 02/3/2010, and partial support by the Bulgarian NSF grant DO 02-85.
\end{acknowledgements}

\end{document}